\documentclass[%
 reprint,
 superscriptaddress,
showpacs,preprintnumbers,
 amsmath,amssymb,
 aps,
prb,
]{revtex4-1}

\usepackage{graphicx}
\usepackage{dcolumn}
\usepackage{bm}
\usepackage{hyperref}

\begin{document}

\preprint{APS/123-QED}

\title{Temperature-dependent magnetospectroscopy of HgTe quantum wells}

\author{A.~V.~Ikonnikov}
\email{antikon@ipmras.ru}
\affiliation{Institute for Physics of Microstructures RAS, GSP-105, 603950, Nizhny Novgorod, Russia}
\affiliation{Lobachevsky State University of Nizhny Novgorod, 603950, Nizhny Novgorod, Russia}

\author{S.~S.~Krishtopenko}
\affiliation{Institute for Physics of Microstructures RAS, GSP-105, 603950, Nizhny Novgorod, Russia}
\affiliation{Laboratoire Charles Coulomb (L2C), UMR CNRS 5221, GIS-TERALAB, Universite Montpellier II, 34095, Montpellier, France}

\author{O.~Drachenko}
\affiliation{Laboratoire National des Champs Magnetiques Intenses (LNCMI-T), CNRS, UPR 3228 Universite de Toulouse, 143 avenue de Rangueil, F-31400 Toulouse, France}

\author{M.~Goiran}
\affiliation{Laboratoire National des Champs Magnetiques Intenses (LNCMI-T), CNRS, UPR 3228 Universite de Toulouse, 143 avenue de Rangueil, F-31400 Toulouse, France}

\author{M.~S.~Zholudev}
\affiliation{Institute for Physics of Microstructures RAS, GSP-105, 603950, Nizhny Novgorod, Russia}

\author{V.~V.~Platonov}
\author{Yu.~B.~Kudasov}
\author{A. S. Korshunov}
\author{D.~A.~Maslov}
\affiliation{Sarov State Institute of Physics and Technology, National Research Nuclear University MEPhI, 607188, Sarov, Russia}
\affiliation{Scientific and Technical Center of Physics, Russian Federal Nuclear Center – All-Russian Research Institute of Experimental Physics, 607188, Sarov, Russia}

\author{I.~V.~Makarov}
\author{O.~M.~Surdin}
\author{A.~V.~Philippov}
\affiliation{Scientific and Technical Center of Physics, Russian Federal Nuclear Center – All-Russian Research Institute of Experimental Physics, 607188, Sarov, Russia}

\author{M.~Marcinkiewicz}
\author{S.~Ruffenach}
\author{F.~Teppe}
\author{W.~Knap}
\affiliation{Laboratoire Charles Coulomb (L2C), UMR CNRS 5221, GIS-TERALAB, Universite Montpellier II, 34095, Montpellier, France}

\author{N.~N.~Mikhailov}
\author{S.~A.~Dvoretsky}

\affiliation{Institute of Semiconductor Physics, Siberian Branch RAS, 630090, Novosibirsk, Russia}

\author{V.~I.~Gavrilenko}
\affiliation{Institute for Physics of Microstructures RAS, GSP-105, 603950, Nizhny Novgorod, Russia}
\affiliation{Lobachevsky State University of Nizhny Novgorod, 603950, Nizhny Novgorod, Russia}

\date{\today}

\begin{abstract}
We report on magnetospectroscopy of HgTe quantum wells in magnetic fields up to 45~T in temperature range from 4.2~K up to 185~K. We observe intra- and inter-band transitions from zero-mode Landau levels, which split from the bottom conduction and upper valence subbands, and merge under the applied magnetic field. To describe experimental results, realistic temperature-dependent calculations of Landau levels have been performed. We show that although our samples are topological insulators at low temperatures only, the signature of such phase persists in optical transitions at high temperatures and high magnetic fields. Our results demonstrate that temperature-dependent magnetospectroscopy is a powerful tool to discriminate trivial and topological insulator phases in HgTe quantum wells.
\end{abstract}

\pacs{76.40.+b, 78.66.-w, 81.07.St, 71.70.Di, 71.20.-b}
\maketitle



Topological insulator (TI) is the quantum state of matter \cite{a1,a2,a3,a4,a5}, characterized by an energy gap in the bulk and conductive boundary (surface) states with linear dispersion law. These states are protected from impurity scattering and electron-electron interactions by time-reversal symmetry \cite{a3,a4,a5,a6,a7}. It means that the electrons in these states can move along the edge of the ultra-thin film or the surface of bulk material without energy loss that can be exploited in ultra-fast and low power consumption electronics. These materials may be used to create groundbreaking electronic devices \cite{a8,a9}, as well as to reveal unusual physical effects occurring at the interfaces of the TI/superconductor and TI/ferromagnetic \cite{a10,a11}.

The first two-dimensional (2D) system, in which TI state was predicted \cite{a1} and then experimentally observed \cite{a2}, were HgTe/CdHgTe quantum wells (QWs). The TI state originates from the inverted band structure in wide HgTe QWs. Specifically, as the QW thickness $d$ increases, the lowest 2D subband in the conduction band formed by $(|\Gamma_6, \pm1/2\rangle)$ states and light-hole $(|\Gamma_8, \pm1/2\rangle)$ states and defined as $E1$ subband, crosses at $d = d_c$ the top subband in the valence band, formed by heavy-hole $(|\Gamma_8, \pm3/2\rangle)$ states, defined as $H1$ subband \cite{a1}. The inverted alignment of electronic states in wide QWs ($d > d_c$) induces spin-polarized helical states at the sample edges \cite{a1}. The existence of such edge states in HgTe QWs has been confirmed experimentally \cite{a2, a12, a13}. At critical QW thickness $d_c$, corresponding to quantum phase transition between conventional semiconductor and TI, the energy structure in the vicinity of band crossing mimics massless Dirac fermions at the $\Gamma$ point \cite{a14, a15, a16}.

The inherent property of inverted HgTe QWs is their characteristic behavior under applied magnetic field $B$, i.e., the crossing of particular zero-mode Landau levels (LLs), arising at critical magnetic field $B_c$ \cite{a2, a12, a13}. Below this field, the lowest zero-mode LL has electron-like character, although it origins from the valence band. This LL tends toward high energies with increasing of magnetic field. The second zero-mode LL, which arises from conduction band at $B < B_c$, has the heavy-hole-like character and decreases with magnetic field. In this situation, counter propagating spin-polarized states also still exist \cite{a2}, although, owing to the presence of magnetic field and breaking of time-reversal symmetry, these states are not robustly protected. For $B > B_c$, the band structure becomes normal and only trivial quantum Hall states can be found. Recently, increasing of the critical QW thickness with temperature has been shown \cite{a17}. The latter indicates that TI state is destroyed if temperature increases. Later, the temperature-induced phase transition between inverted and normal band structure has been confirmed by magnetotransport studies \cite{a18}.

Up to now, to discriminate between trivial ($d < d_c$) and topological insulators ($d > d_c$) in HgTe QWs, especially close to the critical width, one has to performed detailed magnetotransport investigations of gated Hall bars \cite{a12, a13, a18}. In this work, we demonstrate that such difference can also be made by magnetooptical measurements at different temperatures on non-processed samples. We note that previous magnetooptical studies \cite{a15, a16, a19, a20, a21, a22, a26, z1} of HgTe QWs have been performed at low temperatures only.

We perform Landau level magnetospectroscopy in wide temperature range up to 185~K of two HgTe/Cd$_x$Hg$_{1-x}$Te QWs, which are in TI regime at low temperatures. Our samples were grown by molecular beam epitaxy (MBE) on semi-insulating GaAs(013) substrates \cite{a27}. CdTe buffer, $\sim$40~nm lower Cd$_x$Hg$_{1-x}$Te barrier, HgTe QW, and $\sim$40~nm Cd$_x$Hg$_{1-x}$Te top barrier were grown one by one. 40~nm CdTe cap layer was also grown above the structure. The Cd content $x$ in the barriers and QW width are given in the Table~\ref{tab:1}. The barriers of both samples were selectively doped with indium (from both sides of QW), that resulted in formation of 2D electron gas in the QW with a concentration of several units of $10^{11}$~cm$^{−2}$ at low temperatures. Typical mobility values at low temperatures are about 5${\times}10^4$~cm$^2$/V$\cdot$s. Both samples have inverted band structure at low temperatures.


\begin{table}
\caption{\label{tab:1}Parameters of HgTe/Cd$_x$Hg$_{1-x}$Te QWs at $T = 4.2$~K.}
\begin{ruledtabular}
\begin{tabular}{lccc}
Sample&QW width (nm)&$x_{bar}$ (\%)&$n_s$ ($10^{11}$ cm$^{-2}$)\\
\hline
1 (091223-1)&8&62&1.6\\
2 (091222-1)&8&70&3.2\\
\end{tabular}
\end{ruledtabular}
\end{table}

Magnetooptical experiments were performed in pulsed magnetic fields up to 45~T in Laboratoire National des Champs Magnetiques Intenses in Toulouse (LNCMI-T) and in Sarov State Institute of Physics and Technology (SSIPT). The pulse duration in LNCMI-T experiments was about 800~ms, while pulse duration in SSIPT did not exceed 25~ms. Solenoids were immersed into a liquid nitrogen Dewar. Each setup had its own peculiarities.

In LNCMI-T, the liquid helium cryostat was placed inside the solenoid. The variable temperature probe with a sample, quantum cascade laser (QCL) emitting at wavelength of 14.8~$\mu$m and blocked impurity band Si detectors were inserted into the cryostat. It allowed to perform magnetooptical measurements in the temperature range from 4.2~K to 30~K. Details about this setup are given in Ref.~\onlinecite{a28}.

In SSIPT, the samples were mounted on the cold finger in the vacuum chamber of liquid nitrogen cryostat accompanied by temperature sensor and magnetic field induction sensors. The temperature was varied in the range of 77-185~K \cite{a29}. As a radiation source, we used CO$_2$ laser ($\lambda = 10.6$~$\mu$m), while detector was HgCdTe photodiode operating at liquid nitrogen temperature. Magnetooptical measurements were made in the Faraday configuration. Additionally, to determine electron concentration and LL filling factors $\nu$ at different temperatures, we also measured magnetoresistance in two-terminal geometry.

\begin{figure*}
\includegraphics[width=0.33\textwidth]{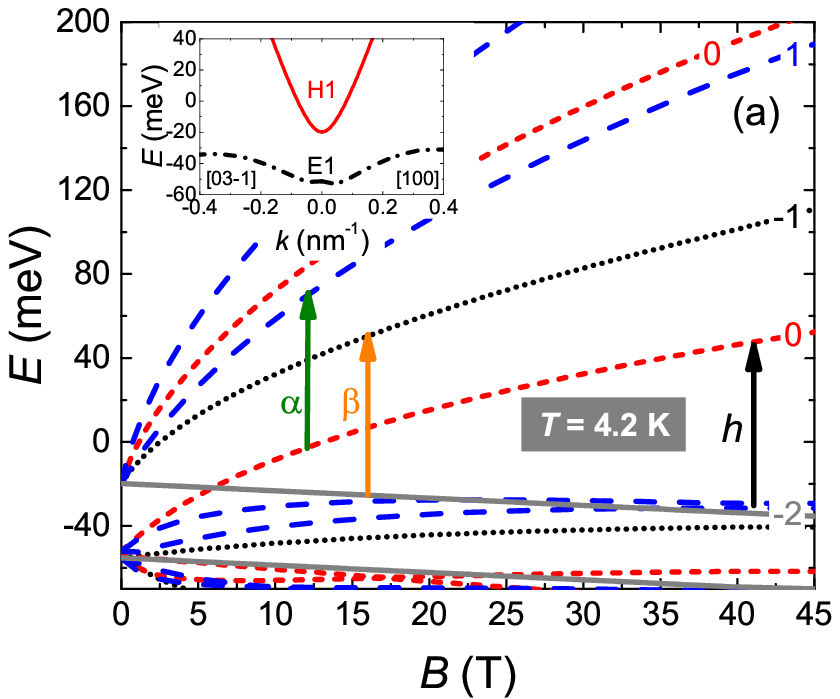}
\includegraphics[width=0.33\textwidth]{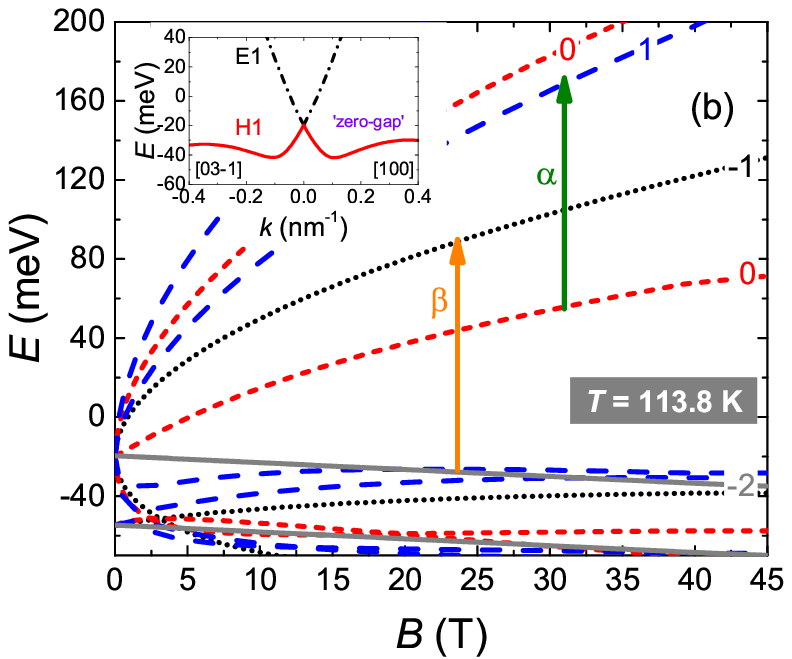}
\includegraphics[width=0.33\textwidth]{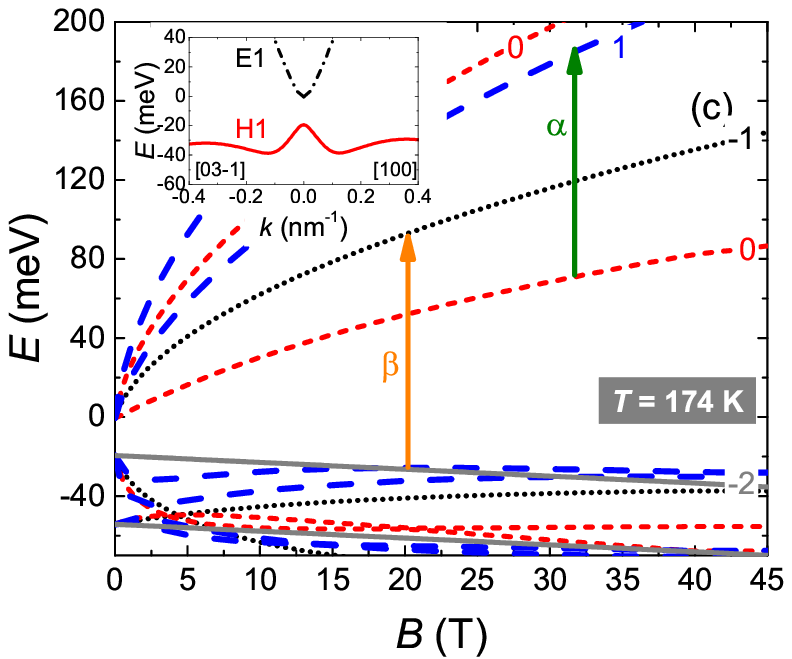}
\caption{\label{fig:1} (Color online) Landau levels (in the axial approximation) and band structure at $B = 0$ (insets) for $k\| [100]$ and $k \| [03-1]$ for the sample~1 at different temperatures: (a) $T = 4.2$~K, the band structure is inverted with an indirect band gap; (b) $T = 113.8~K$, a gapless state (the inset shows a Dirac cone in the vicinity of the $\Gamma$ point); (c) $T = 174$~K, the band structure is normal with a direct band gap, conduction subband $E1$ has an electron-like character, the valence subband is formed by ‘hole-like’ level $H1$}
\end{figure*}

To interpret the experimental results, we performed temperature-dependent band structure and LLs calculations based on the 8-band \textbf{k${\cdot}$p} Hamiltonian for (013)-oriented heterostructures (see, e.g. Refs.~\onlinecite{a15, u30}) with material parameters taken from Ref.~\onlinecite{a18}. In the model, we also take into account a tensile strain in the layers arising due to the mismatch of lattice constants in CdTe buffer, HgTe QW and Cd$_x$Hg$_{1-x}$Te barriers. The calculations were performed by expanding the envelope wave functions in the basis set of plane waves and by numerical solution of the eigenvalue problem. The energies of LLs were found within so-called axial approximation \cite{a15, u30}, while for the calculations of dispersion curves non-axial terms were held. In our approach we takes into account the temperature dependence of the bandgap, valence band offset, change in the lattice constants of the layers and the elastic constants $C_{11}$, $C_{12}$ and $C_{44}$ (bulk modulus) with the temperature \cite{a18, b31}.

Fig.~\ref{fig:1} provides a LL fan chart and dispersion curves for the sample~1 at three different temperatures. At low temperatures, the sample is in 2D TI phase and conduction band is formed by the top 'hole-like' subband $H1$. The lowest LL in the conduction band, labeled by $n = -2$, has a purely heavy-hole character $(|\Gamma_8, -3/2\rangle)$ and its energy decreases linearly with magnetic field $B$. For the notations of LLs, see Refs.~\onlinecite{a15, a19}. In contrast, the top LL in the valence band $n = 0$ goes up in energy with a magnetic field. These two LLs represent, so-called zero-mode LLs, mentioned above, which are identified within a simplified approach of the Dirac type Hamiltonian \cite{a1}. In calculations of LLs in our samples, we applied a general scheme of the 8-band \textbf{k${\cdot}$p} Hamiltonian but neglected the bulk inversion asymmetry (BIA) effect \cite{a22}. Such an approximation implies that, for any HgTe QW in the inverted regime, the two zero-mode LLs simply cross each other at given magnetic field $B_c$. These characteristic levels and their crossing can be easily recognized in Fig.~\ref{fig:1} for the sample~1. It is seen that in magnetic fields above 6.4 T (at $T = 4.2$~K), the sample~1 has normal band structure, and the level $n = 0$ becomes actually the lowest LL in the conduction band. In the models with BIA \cite{a22}, the level crossing between zero-mode LLs at $B_c$ can be avoided. The latter gives rise to specific behaviour of magnetooptical transitions from the zero-mode LL in the vicinity of critical magnetic field \cite{a15, a19, a22}. If magnetic field exceeds $B_c$ or is significantly lower than the critical field, effect of BIA is negligible small. Further, BIA effect is neglected.

As it is seen in Fig.~\ref{fig:1}, the band gap and $B_c$ are getting smaller with temperature $T$. At critical temperature $T_c = 113.8$~K for the sample~1, the band gap vanishes, and the band structure mimics dispersion of massless Dirac fermions. The further increasing of $T$ opens the band gap and makes HgTe QW a conventional semiconductor with normal band ordering, in which conduction band has ‘electron-like’ character, while the valence band at the $\Gamma$ point is formed by heavy-hole states. Thus, a temperature increase results in a qualitative transformation of the inverted band structure into the normal one.

\begin{figure}
\includegraphics[width=\columnwidth]{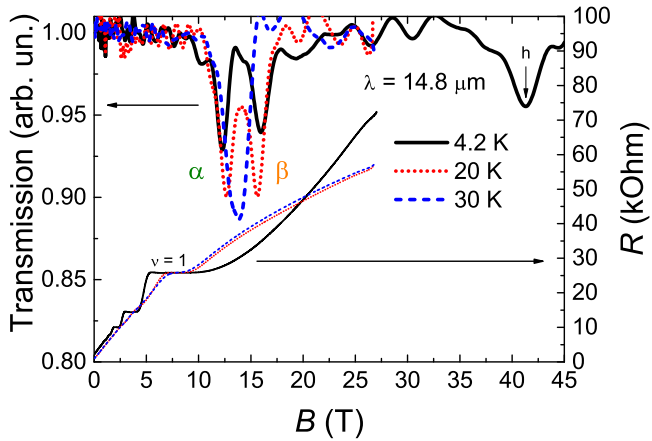}
\caption{\label{fig:2} (Color online) Magnetoresistance and magnetoabsorption spectra for the sample~1, obtained at different temperatures (solid lines -- 4.2~K, dotted -- 20~K, dashed -- 30~K) using 14.8~$\mu$m QCL.}
\end{figure}

\begin{figure}
\includegraphics[width=\columnwidth]{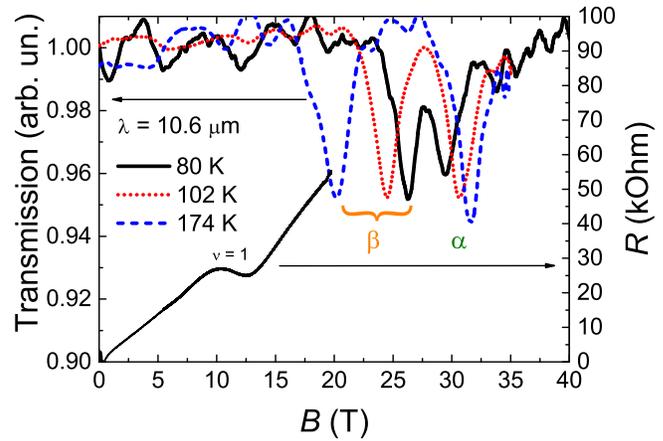}
\caption{\label{fig:3} (Color online) Magnetoresistance and magnetoabsorption spectra for the sample~1, obtained at different temperatures (solid lines -- 80~K, dotted -- 102~K, dashed -- 174~K) using CO$_2$ laser emitting at 10.6~$\mu$m.}
\end{figure}

Fig.~\ref{fig:2} shows magnetic field dependences of two-terminal magnetoresistance and transmission measured in the sample~1 by using 14.8~$\mu$m QCL at three different temperatures. The absorption spectrum exhibits three lines denoted by $\alpha$, $\beta$ and $h$. As it can be seen from magnetoresistance, the quantum Hall plateau, corresponding to LL filling factor $\nu = 1$, occurs within the interval of the magnetic field of 5 to 10~T, and all the lines are observed at higher magnetic fields, for which $\nu$ is less than unity, i.e. in the ultra-quantum limit. In this case, the Fermi level lies at the zero-mode LL with $n = 0$. The selection rules for electric-dipole transitions in the axial approximation allow electron excitation between LLs whose numbers differ by one. Therefore, the observed absorption lines correspond to $0 \rightarrow 1$ electron transition ($\alpha$ line) between the partial occupied upper zero-model LL and high-lying LL with $n = 1$ and also an electron excitation from the lower zero-model LL with $n = -2$ into high-lying empty LL with $n = -1$ (see Fig.~\ref{fig:1}a). The latter is called $\beta$ line \cite{a19, a20, a21, a22, z1}. In addition to $\alpha$ and $\beta$ lines, an intra-band electron transition from LL with $n = –1$ in the valence band into partially filled zero-mode LL with $n = 0$ is clearly observed. This line is designated as $h$ (see Fig.~\ref{fig:1}, cf. \onlinecite{a21}).

\begin{figure}
\includegraphics[width=\columnwidth]{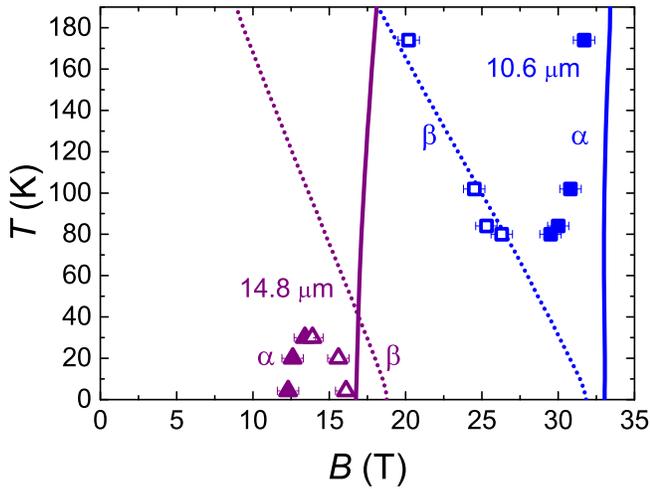}
\caption{\label{fig:4} (Color online) Temperature dependence of magnetoabsorption line positions for two wavelengths: 10.6~$\mu$m (squares), and 14.8~$\mu$m (triangles) for the sample~1. The curves stand for the calculation results; symbols are experimental data, extracted from the absorption maxima. Solid curves and symbols correspond to $\alpha$ line ($0 \rightarrow 1$ transition), dotted curves and open symbols are for the $\beta$ line ($-2 \rightarrow –1$ transition).}
\end{figure}

If temperature increases, the $\alpha$ and $\beta$ lines merge (Fig.~\ref{fig:2}a). However, at high temperature, evolution of $\alpha$ and $\beta$ lines has a different behaviours. Fig.~\ref{fig:3} shows magnetospectroscopy results for the sample~1 obtained with CO$_2$ laser at 10.6~$\mu$m at different temperatures $T \geq 80$~K. As in Fig.~\ref{fig:2}, all the absorption lines are observed in magnetic fields exceeding the range for of fundamental quantum Hall plateau $\nu=1$. The latter is shifted towards higher magnetic fields due to increased electron concentration if compared with the plateau at 4.2~K. It is seen that the $\alpha$ line is observed in higher fields than the $\beta$ line (cf. Fig.~\ref{fig:1}(b, c)). The temperature increase results in the divergence of the lines: the $\beta$ line rapidly tends to low magnetic fields, while the $\alpha$ line slowly shifts toward high field region. In order to get a detailed picture of temperature-dependent magnetospectroscopy, we plot the resonant magnetic fields, corresponding to the absorption maxima, as a function of temperature for two wavelengths used in our experiments (Fig. ~\ref{fig:4}). It is seen that the resonant fields for $0 \rightarrow 1$ transition ($\alpha$ line, closed symbols) weakly depend on $T$ at high temperatures. In contrast, resonant fields for $-2 \rightarrow -1$ transition ($\beta$ line, open symbols) strongly depend on temperature, shifting toward low fields with temperature.

\begin{figure}
\includegraphics[width=\columnwidth]{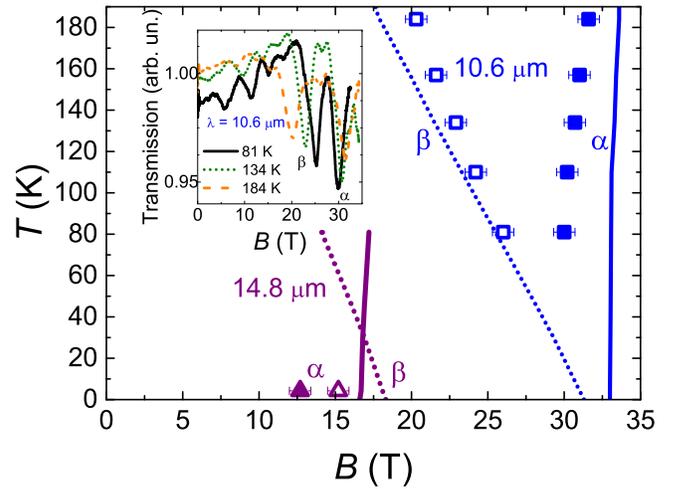}
\caption{\label{fig:5} (Color online) Temperature dependence of magnetoabsorption line positions for two wavelengths: 10.6~$\mu$m (squares), and 14.8~$\mu$m (triangles) for the sample~2. Lines stand for the calculation results, symbols are experimental data. Solid lines and symbols correspond to the line $\alpha$ ($0 \rightarrow 1$ transition), dotted lines and open symbols correspond to the line $\beta$ ($-2 \rightarrow -1$ transition). The inset shows typical magnetoabsorption spectra, obtained at different temperatures (solid lines – 81~K, dotted – 134~K, dashed – 184~K) in pulsed magnetic fields with CO$_2$ laser ($\lambda = 10.6$~$\mu$μm).}
\end{figure}

The sample~2 has the same QW width of 8~nm but higher cadmium content in the barriers than the one for the sample~1. Therefore, the calculated critical temperature $T_c$ and the temperature for the $\alpha$ and $\beta$ line merging in the sample~2 are lower than for the sample~1 (cf. Fig.~\ref{fig:4}, \ref{fig:5}, $\lambda = 14.8$~$\mu$m). Temperature-dependent measurements for the sample~2 were carried out with CO$_2$ laser only; those with QCL were performed at $T = 4.2$~K (Fig.~\ref{fig:5}). Just as in the sample~1, the observed line positions correspond to the quantum limit $\nu \leq 1$. One can see that at $\lambda = 14.8$~$\mu$m, $T = 4.2$~K, the $\alpha$ and $\beta$ lines, indeed, are a bit closer to each other than in the sample~1. At high temperatures $T \geq 80$~K, the experimental results at wavelength of 10.6~$\mu$m are very similar to those for the sample~1. As easy to see from Fig.~\ref{fig:4}, \ref{fig:5}, there is a good qualitative agreement between experimental data and theoretical results. The slopes of calculated temperature dependences of line positions are close to those observed experimentally. This indicates that the selected temperature dependences of the band parameters\cite{a18} used in the 8-band \textbf{k${\cdot}$p} Hamiltonian are adequate.

\begin{figure*}
\includegraphics[width=0.5\textwidth]{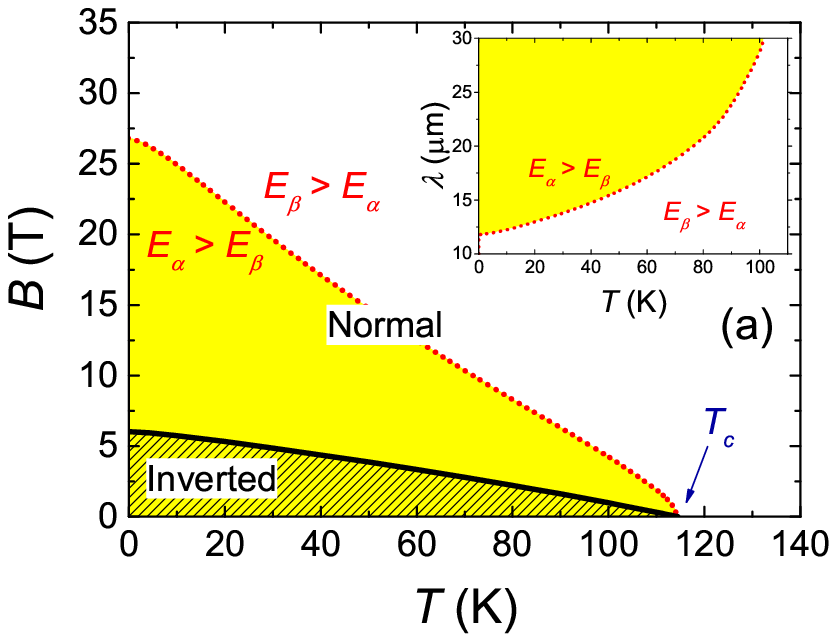}
\includegraphics[width=0.5\textwidth]{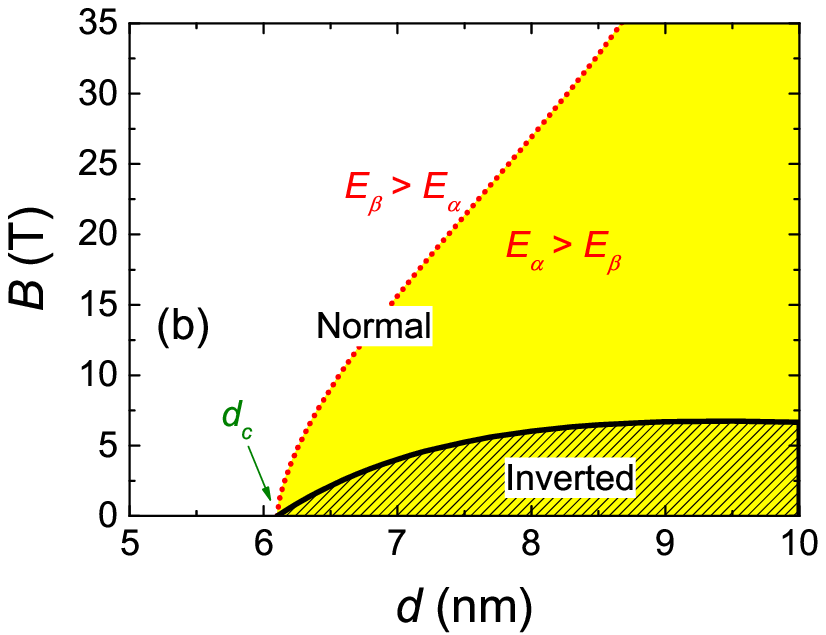}
\caption{\label{fig:6} (Color online)
Phase diagrams for HgTe/Cd$_{0.62}$Hg$_{0.38}$Te~QW at different values of temperature (a, QW width $d = 8$~nm) and QW width $d$ (b, $T = 0$). Striped regions correspond to the inverted band structure. The red dotted curves conform to the magnetic fields in the which $\alpha$ and $\beta$  lines merge ($E_\alpha = E_\beta$). The inset on the left figure shows the wavelength, at which positions of the $\alpha$ and $\beta$ lines coincide, as a function of temperature.}
\end{figure*}

Different behavior of the $\alpha$ and $\beta$ lines (Fig.~\ref{fig:4}, ~\ref{fig:5}) results from the temperature effect on the band structure in HgTe QWs. Roughly speaking, at low temperatures, the $\alpha$ line corresponds to the inter-band transition, while at high temperatures it becomes the intra-band transition. Since the inverted gap is closing up to $T_c$, at fixed magnetic field, the energy of the $\alpha$ transition should decrease with the temperature) and vice versa at fixed excitation energy, the resonant magnetic field should dramatically increase with temperature. On the contrary, the $\beta$ line at low temperature corresponds to the intra-band transition, while at high temperatures it results from the inter-band excitation. Therefore, if magnetic field is fixed, the energy of the $\beta$ line should increase with temperature due to the band gap opening over $T_c$. Therefore, at fixed excitation energy the resonant magnetic field for the $\beta$ line should decrease if the temperature increases.

Fig.~\ref{fig:6}a provides the phase diagram for the sample~1 at different values of magnetic field and temperature. The black solid curve shows dependence of $B_c$ on temperature. It confines the striped region, corresponding to the inverted band structure. Above this curves, the sample~1 has normal band structure. We note that at critical temperature $T_c = 113.8$~K, $B_c = 0$; the latter corresponds to the gapless state with a Dirac cone in the vicinity of the Γ point, shown in Fig.~\ref{fig:1}b. We also plot in Fig.~\ref{fig:6}a the temperature dependence of specific magnetic field, at which the $\alpha$ and $\beta$ lines coincide. This is given by the red dotted curve. Below this curve, resonant energy of the $\alpha$ line exceeds that of the $\beta$ line. If the temperature tends to $T_c$, energies of $\alpha$ and $\beta$ transitions coincide at $B \rightarrow 0$. As it can be demonstrated analytically, for an example, be means of simplified approach\cite{a1}, the latter results from arising of the Dirac cone in the vicinity of the $\Gamma$ point.

The crossing of $\alpha$ and $\beta$ lines with the temperature increasing is the signature of the inverted band structure at low temperatures. It is easy to verify within the Dirac type Hamiltonian\cite{a1}, that such crossing in given magnetic field is related with negative values of the mass parameter $M$. The latter is absent for the positive values of $M$, i.e. for the trivial insulator phase. In Fig.~\ref{fig:6}b, we also provide the phase diagram for HgTe/Cd$_{0.62}$Hg$_{0.38}$Te QW at zero temperature as a function of QW width $d$. One can see that the merging of the $\alpha$ and $\beta$ lines with energies $E_{\alpha}$ = $E_{\beta}$ (with increasing of magnetic field) takes place at $d > d_c$ only. The latter corresponds to the 2D TI phase in zero magnetic field.

Let us now explain a different temperature evolution for the $\alpha$ and $\beta$  absorption lines at the wavelength of 10.6~$\mu$m and 14.8~$\mu$m. To probe the crossing of $\alpha$ and $\beta$ transitions, in addition to the range of temperatures and magnetic fields, one needs also to choose a proper frequency range. The inset in Fig.~\ref{fig:6}a shows theoretical temperature dependence of the wavelength for the crossing of $\alpha$ and $\beta$ transitions in the sample~1. It is seen that, at given temperature, there is a short wavelength limit for probing coincidence between the $\alpha$ and $\beta$ absorption lines. Indeed, the wavelength of CO$_2$ laser $\lambda = 10.6$~$\mu$m does not allowed to probe a merging of the lines; the splitting between the lines increases with temperature (see Fig.~\ref{fig:2}). The picture is changes drastically for $\lambda = 14.8$~$\mu$m. It is seen that the $\alpha$ and $\beta$ transitions have the same energies for $T \approx 40$~K. In the lower temperature range, the $\alpha$ and $\beta$ absorption lines merge if temperature increases (see Fig.~\ref{fig:2}).

In summary, we have performed the temperature-dependent magnetospectroscopy study in pulsed magnetic fields up to 45~T of HgTe/CdHgTe QWs with inverted (at low temperatures) band structure by means of monochromatic sources. At low excitation energies, we have discovered a temperature-induced merging of the absorption lines, corresponding to the transitions from the zero-mode LLs. Realistic temperature-dependent calculations of LLs, based on the 8-band \textbf{k${\cdot}$p} Hamiltonian, allows us to interpret such behaviour of the observed transition as a residual signature of low-temperature TI phase, which fingerprint persists at high temperatures and high magnetic fields. Our results demonstrate that temperature-dependent magnetospectroscopy can be used as a tool to probe a difference between trivial and topological insulator phases in HgTe quantum wells.

This work was supported by the Russian Scientific Foundation (Grant No 16-12-10317), CNRS through LIA TeraMIR project, by Languedoc-Roussillon region via the “Terapole Gepeto Platform”. The authors thank Vladimir Aleshkin for helpful discussions and comments on this work. S.S. Krishtopenko also acknowledges the non-profit Dynasty foundation for financial support.

\nocite{*}

\bibliography{ikonnikov3}

\end{document}